# Analyzing the Efficiency of $M_n$-$(C_2H_4)$ (M = Sc, Ti, Fe, Ni; n = 1, 2) Complexes as Effective Hydrogen Storage Materials

**Arindam Chakraborty,[a] Santanab Giri[a] and Pratim Kumar Chattaraj*[a]**

Department of Chemistry and Center for Theoretical Studies
Indian Institute of Technology Kharagpur
Kharagpur- 721302

E-mail: pkc@chem.iitkgp.ernet.in

**Abstract:**

Hydrogen trapping ability of various metal – ethylene complexes has been studied at the B3LYP and MP2 level of theory using the 6-311+G(d,p) basis set. Different global and local reactivity descriptors and the associated electronic structure principles provide important insights into the associated interactions. There exist two distinct classes of bonding patterns, viz., a Kubas-type interaction between the metal and the H2 molecule behaving as a η2-ligand and an electrostatic interaction between the metal and the atomic hydrogens.

**Introduction**

Hydrogen, the third most abundant element on Earth and perhaps the most abundant element on the Universe is being conceived as a very promising alternative energy carrier as well as a future fuel reserve for the transportation industry. In its molecular form hydrogen can be used directly as a fuel to drive a vehicle, to heat water or indirectly to produce electricity for industrial, transport and domestic use. The superiority of hydrogen over the so-called fossil fuels lies in its sheer "cleanliness", a unique fuel that is totally non-polluting and upon use produces water as a harmless by-product. Unlike petroleum, hydrogen can be easily generated from renewable energy resources which further eliminate the production of oxides of nitrogen and sulfur, greenhouse gases like carbon dioxide and methane as by-products thereby eradicating further scopes of environmental pollution. But the storage of gaseous hydrogen in a practical sense creates difficulties as the materials that can trap the same in large gravimetric and volumetric quantities are really numbered. In liquid form, hydrogen can only be stored under cryogenic temperatures which can never be a good option for day by day use. But scientists nevertheless have been somewhat successful in designing novel materials that can store hydrogen at ambient conditions.[1-3] Numerous other materials, like aluminum nitride (AlN) nanostructures[4], transition-metal doped boron nitride (BN) systems[5], alkali-metal doped benzenoid[6] and fullerene clusters[7], bare as well as light metal and transition-metal coated boron buckyballs, $B_{80}$[8], and magnesium clusters[9] have been confirmed both experimentally and theoretically to serve as potential hydrogen-storage materials. Further, based on a theoretical study invoking the metastability of hydrogen-stacked $Mg_n$ clusters[10], Chattaraj et al[11] have very recently demonstrated that a host of small to medium metal cluster moieties involving $Li_3^+$, $Na_3^+$, $Mg_n$ and $Ca_n$ (n = 8-10) cages have got a fair capability of trapping hydrogen in both atomic and molecular forms. The stability of the aforesaid all-metal systems has been attributed to the existence of an aromaticity criterion in the metallic rings which was assessed through the rationale of nucleus independent chemical shift (NICS)[12]. Yildirim et al[13-18] along with a few other research groups[19,20] have been quite successful in establishing that the C=C bond in an ethylene molecule, $C_2H_4$, like that of fullerenes and other carbon-based nanostructures can form stable complexes with a transition metal, Titanium (Ti) and the resulting $Ti_n$–$C_2H_4$ (n = 1, 2) complex in turn can bind up to ten $H_2$ molecules efficiently. The interaction between the hydrogen molecules and the transition metal is worth-mentioning as it is somewhat intermediate between physical and chemical adsorption but, can best be explained from the Kubas model[21] of hydrogen binding: ($\eta^2$-$H_2$) – metal interaction.

In this article we have made an attempt to study the binding of hydrogen with a host of transition metal (M) – ethylene ($C_2H_4$) complexes [$M_n$-($C_2H_4$) (M = Sc, Ti, Fe, Ni; n = 1, 2)] on the basic premise of conceptual density functional theory (CDFT)[22-25] and its various allied global reactivity descriptors like electronegativity[26-28] ($\chi$), hardness[29-31] ($\eta$), electrophilicity[32-34] ($\omega$) and the local variants like atomic charges[35] ($Q_k$) and Fukui functions[36] ($f_k$). The stability of the resulting hydrogen-bound [$M_n$-($C_2H_4$) (M = Sc, Ti, Fe, Ni; n = 1, 2)] complexes may be understood from the corresponding interaction energies ($\Delta E$) and reaction electrophilicities ($\Delta \omega$) of a number of plausible trapping reactions. Further, for most of the hydrogen bound metal – ethylene complexes [$(H_2)_x$ – $M_n$-($C_2H_4$) (M = Sc, Ti, Fe, Ni; n = 1, 2; x = 1-8)], hydrogen prefers mostly to coordinate with the metal center (M) in its molecular (dihydrogen) form. The $H_2$ molecule therefore approximately behaves as a $\eta^2$-type ligand and the complex subsequently attains stability though a Kubas-type[21] interaction between the metal atom and the interacting dihydrogen moieties.

**Theoretical Background**

The thermodynamic stability of molecular systems may be meaningfully justified in a quantitative manner from a careful scrutiny of their chemical hardness ($\eta$) and electrophilicity ($\omega$) values. This has been further validated by the establishment of some associated molecular electronic structure principles like the Principle of Maximum Hardness[37-39] (PMH) together with the Minimum Polarizability Principle[40,41] (MPP) and Minimum Electrophilicity Principle[42,43] (MEP). These electronic structure principles serve as major determinants towards assessing the stability and reactivity trends of chemical systems. For an N-electron system, the electronegativity[26-28] ($\chi$) and hardness[29-31] ($\eta$) can be defined as follows:

$$\chi = -\mu = -\left(\frac{\partial E}{\partial N}\right)_{v(\vec{r})} \qquad (1)$$

$$\eta = \left(\frac{\partial^2 E}{\partial N^2}\right)_{v(\vec{r})} \qquad (2)$$

Here $E$ is the total energy of the $N$-electron system and $\mu$ and $v(\vec{r})$ are its chemical potential and external potential respectively. The electrophilicity[32-34] ($\omega$) is defined as:

$$\omega = \frac{\mu^2}{2\eta} = \frac{\chi^2}{2\eta} \tag{3}$$

A finite difference approximation to Eqs. 1 and 2 can be expressed as:

$$\chi = \frac{I+A}{2} \tag{4}$$

and $\quad \eta = I - A \tag{5}$

where *I* and *A* represent the ionization potential and electron affinity of the system respectively and are computed in terms of the energies of the *N* and $N \pm 1$ electron systems. For an *N*-electron system with energy *E (N)* they may be expressed as follows:

$I = E(N-1) - E(N) \tag{6}$

and $\quad A = E(N) - E(N+1) \tag{7}$

The local reactivity descriptor, Fukui function[36] (FF) measures the change in electron density at a given point when an electron is added to or removed from a system at constant $v(\vec{r})$. It may be written as:

$$f(\vec{r}) = \left(\frac{\partial \rho(\vec{r})}{\partial N}\right)_{v(\vec{r})} = \left(\frac{\delta \mu}{\delta v(\vec{r})}\right)_{N} \tag{8}$$

Condensation of this Fukui function, $f(\vec{r})$ to an individual atomic site k in a molecule gives rise to the following expressions in terms of electron population[44] $q_k$

$f_k^+ = q_k(N+1) - q_k(N)$ for nucleophilic attack $\quad$ (9a)

$f_k^- = q_k(N) - q_k(N-1)$ for electrophilic attack $\quad$ (9b)

$f_k^o = [q_k(N+1) - q_k(N-1)]/2$ for radical attack $\quad$ (9c)

## Computational Details

The geometry optimization and subsequent frequency calculations of the different metal – ethylene complexes $M_n$-$(C_2H_4)$ (M = Sc, Ti, Fe, Ni; n = 1, 2) and their corresponding hydrogen-trapped analogues are carried out at the B3LYP level of theory using the 6-311+G(d,p) basis set with the aid of the GAUSSIAN 03 program package.[45] However, for the di-iron – ethylene complexes [Fe$_2$-(C$_2$H$_4$)] convergence could not be achieved owing to which the aforesaid molecules could not be considered for hydrogen trapping/storage. The number of imaginary frequency (NIMAG) values of all the optimized geometries are zero thereby confirming their existence at the minima on the potential energy surface (PES). Single point calculations are further done to evaluate the energies of the $N \pm 1$ electron systems by adopting the geometries of the corresponding $N$-electron systems optimized at the B3LYP/6-311+G(d) level of theory. The $I$ and $A$ values are calculated using a $\Delta SCF$ technique. The electrophilicity ($\omega$) and hardness ($\eta$) are computed using the eqs. 3 and 5 respectively. A Mulliken population analysis (MPA) scheme is adopted to calculate the atomic charges ($Q_k$) and the corresponding Fukui functions ($f(\vec{r})$) on the metal centers. The frontier molecular orbital pictures are obtained through the GAUSSVIEW 03 package.[45]

**Results and Discussion**

The total energy (E, au) and the important global reactivity descriptors like electronegativity ($\chi$), hardness ($\eta$) and electrophilicity ($\omega$) for all the interacting transition metals, hydrogen (atomic and molecular) and ethylene are shown in table 1. A detailed tabulation of the total energy (E, au) computed at two different level of theory are tabulated in table 2. The above-mentioned global reactivity parameters of the various metal – ethylene complexes M$_n$-(C$_2$H$_4$) (M = Sc, Ti, Fe, Ni; n = 1, 2) and their corresponding hydrogen-trapped analogs [(H$_2$)$_x$ – M$_n$-(C$_2$H$_4$) (M = Sc, Ti, Fe, Ni; n = 1, 2; x = 1-8)] are presented in tables 3, 4, 5 and 6 for the metals Sc, Ti, Fe, and Ni respectively. However, for all the complexed moieties, we have computed the local parameters like atomic charge ($Q_k$) and Fukui functions ($f_k^+$, $f_k^-$) using the Mulliken population analysis (MPA) scheme for the metal site only as it is conceived that the metal center virtually plays the key role towards binding the incoming hydrogen molecules during trapping reactions. A number of plausible hydrogen trapping reactions between the metal – ethylene complexes M$_n$-(C$_2$H$_4$) (M = Sc, Ti, Fe, Ni; n = 1, 2) and the incoming hydrogen molecules for the different transition metals are displayed in tables 7-10. The molecular point groups (PG) of all the complex structures and the corresponding H – H bond

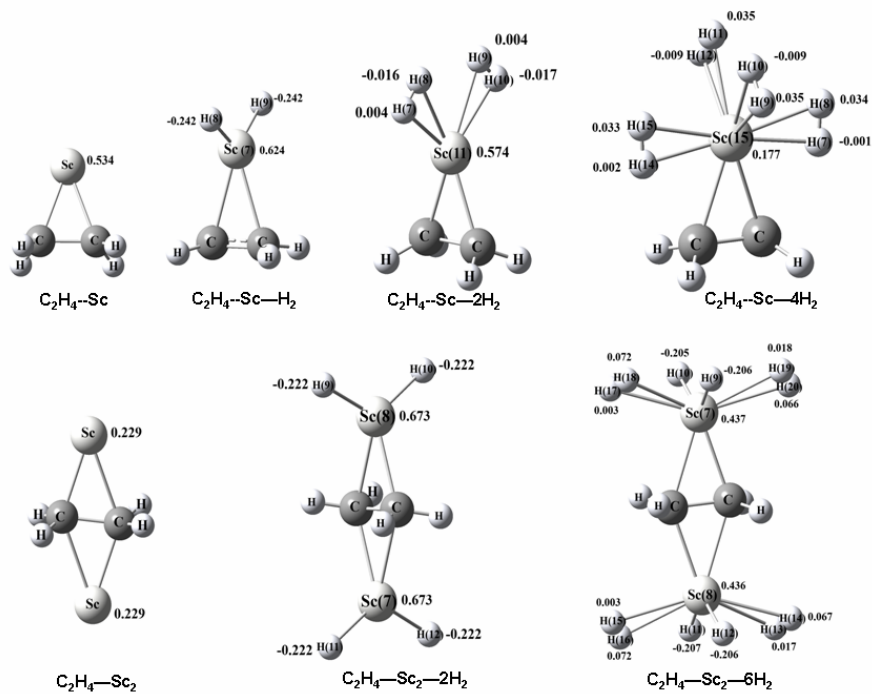

Figure 1. Sc-ethylene complex and its corresponding H2-trapped analogues.

distances (Å) for the H$_2$ molecule(s) trapped on to the metal – ethylene clusters are shown in table 11. The molecular geometries of all the metal – ethylene complexes M$_n$-(C$_2$H$_4$) (M = Sc, Ti, Fe, Ni; n = 1, 2) and their corresponding hydrogen-bound analogues [(H$_2$)$_x$ – M$_n$-(C$_2$H$_4$) (M = Sc, Ti, Fe, Ni; n = 1, 2; x = 1-8)] optimized at the B3LYP/6-311+G(d, p) level of theory are provided in figures 1-4 corresponding to the central metal atom in the order of sequence as Sc, Ti, Fe and Ni. Figures 1 – 4 also depict the atomic charges on

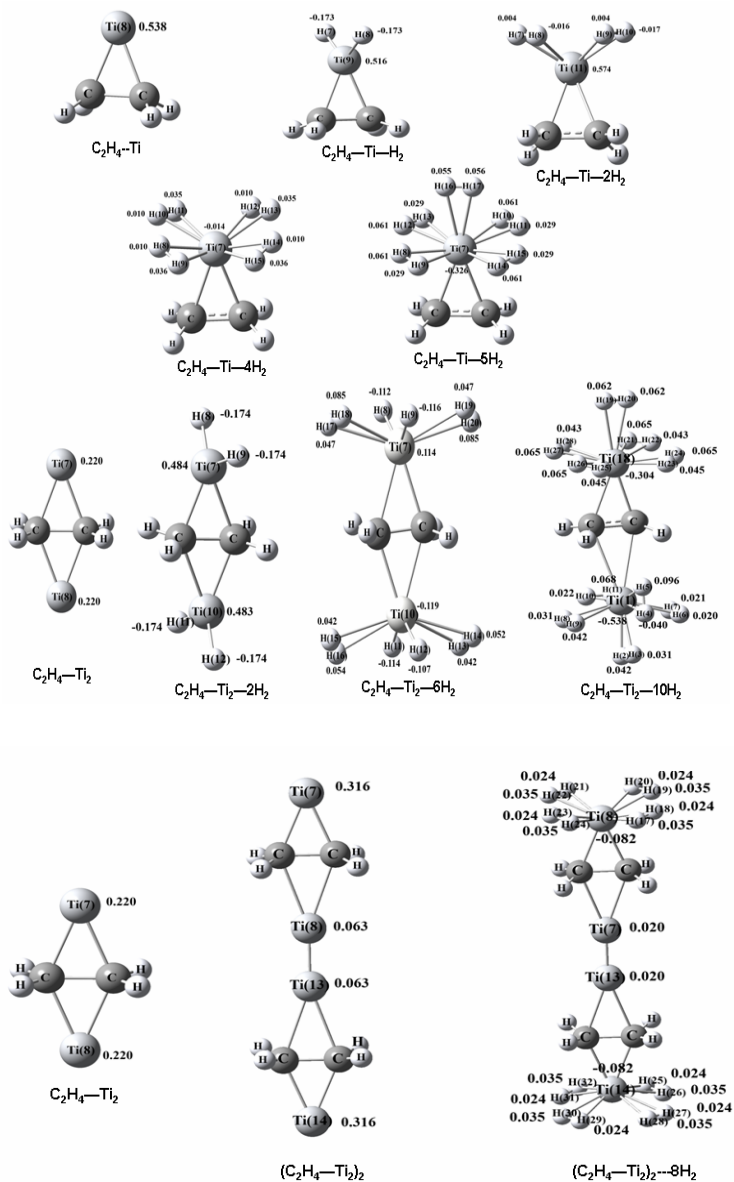

**Figure 2**: Ti-ethylene complex and its corresponding $H_2$-trapped analogues.

the central metal atom as well as the hydrogen centers of the $H_2$ molecules bound to the metal – ethylene moiety either in the split, atomic or closely-bonded molecular form. The important frontier molecular orbitals (FMOs) of all the complex clusters (free and $H_2$-bound) are portrayed in figure 5. A scrutiny of table 1 reveals that the global hardness ($\eta$) of the transition metals approximately decrease from Sc to Fe with a subsequent increase in their respective electrophilicity ($\omega$) values. The energies of the transition metals

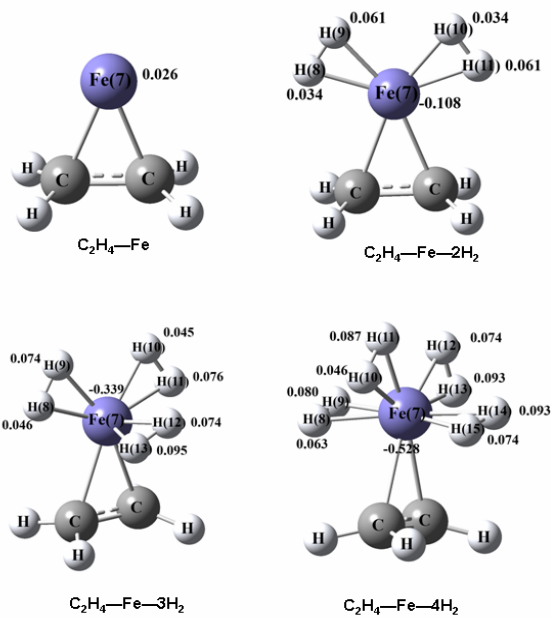

**Figure 3**: Fe-ethylene complex and its corresponding H$_2$-trapped analogues

computed at the B3LYP/6-311+G(d, p) level of theory show an expected increasing trend from Sc to Fe which is quite relevant from their increasing atomic numbers. A detailed analysis of the various metal – ethylene complexes M$_n$-(C$_2$H$_4$) (M = Sc, Ti, Fe, Ni; n = 1, 2) and their corresponding trapping reactions to form the hydrogen-bound analogues can be fruitfully materialized from tables 3 – 10. An analysis of the global and local properties for the Sc – ethylene

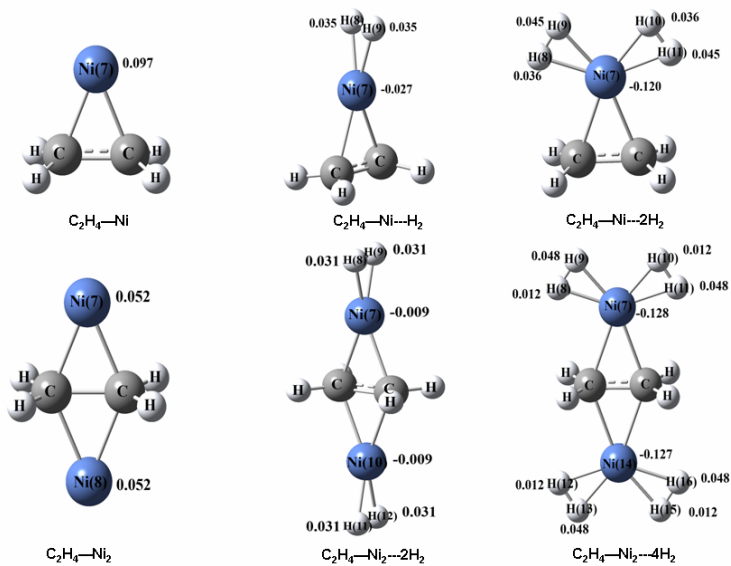

**Figure 4**: Ni-ethylene complex and its corresponding H$_2$-trapped analogues.

complex and hydrogen-trapped analogues as presented in table 3 explains that while the electronegativity ($\chi$) values of the different $Sc_n$ – ethylene (n = 1, 2) complexes do not show any viable change upon transformation from the free, unbound stage to the hydrogen-trapped [$(H_2)_x$ – $Sc_n$-$(C_2H_4)$ (n = 1, 2; x = 1, 2, 4, 6)] form, the hardness ($\eta$) and electrophilicity ($\omega$) exhibit a mixed trend. The $\eta$

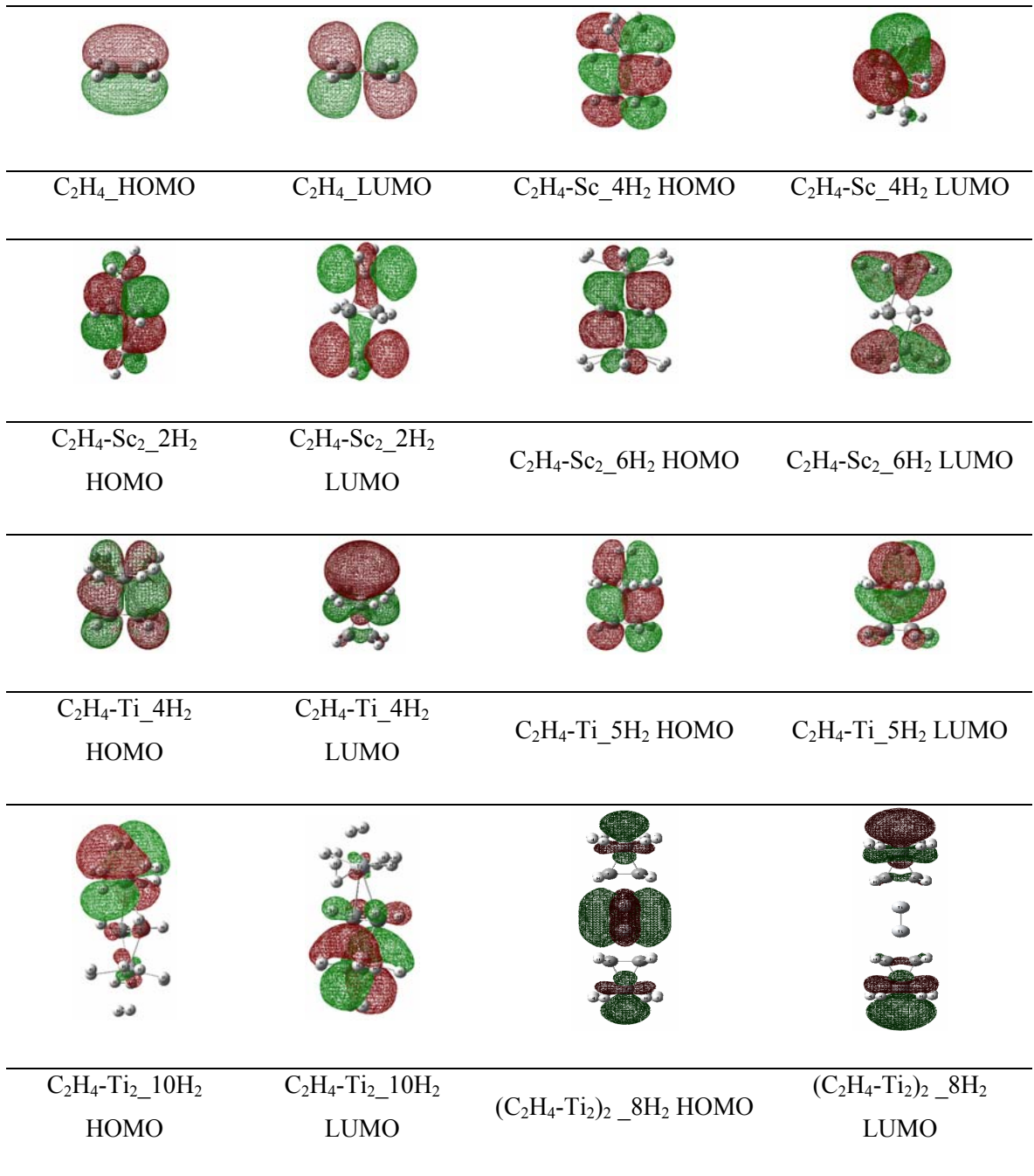

| $C_2H_4$_HOMO | $C_2H_4$_LUMO | $C_2H_4$-Sc_$4H_2$ HOMO | $C_2H_4$-Sc_$4H_2$ LUMO |
| $C_2H_4$-$Sc_2$_$2H_2$ HOMO | $C_2H_4$-$Sc_2$_$2H_2$ LUMO | $C_2H_4$-$Sc_2$_$6H_2$ HOMO | $C_2H_4$-$Sc_2$_$6H_2$ LUMO |
| $C_2H_4$-Ti_$4H_2$ HOMO | $C_2H_4$-Ti_$4H_2$ LUMO | $C_2H_4$-Ti_$5H_2$ HOMO | $C_2H_4$-Ti_$5H_2$ LUMO |
| $C_2H_4$-$Ti_2$_$10H_2$ HOMO | $C_2H_4$-$Ti_2$_$10H_2$ LUMO | $(C_2H_4$-$Ti_2)_2$_$8H_2$ HOMO | $(C_2H_4$-$Ti_2)_2$_$8H_2$ LUMO |

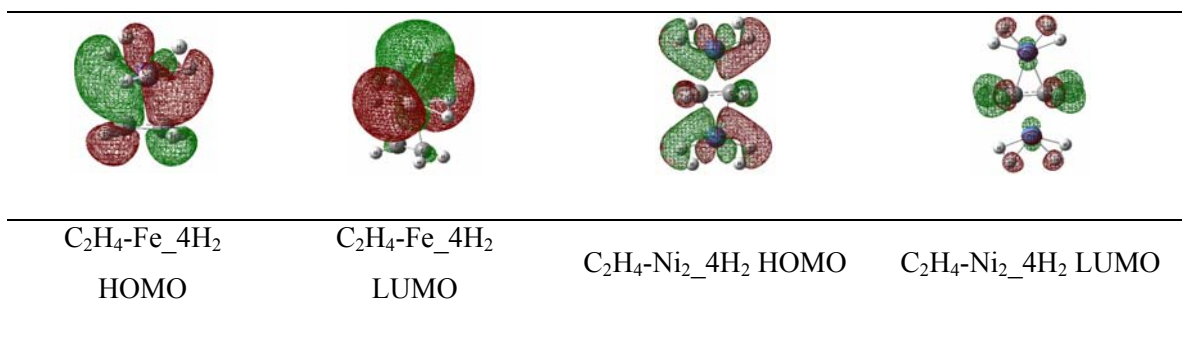

| C$_2$H$_4$-Fe_4H$_2$ HOMO | C$_2$H$_4$-Fe_4H$_2$ LUMO | C$_2$H$_4$-Ni$_2$_4H$_2$ HOMO | C$_2$H$_4$-Ni$_2$_4H$_2$ LUMO |

Figure 5: Some important frontier orbitals (HOMO and LUMO) of H2-trapped metal-ethylene complexes.

**Table 1:** The total energy (E, au), electronegativity ($\chi$, eV), hardness ($\eta$, eV) and electrophilicity ($\omega$, eV) of the interacting transition metals, hydrogen (atomic and molecular) and ethylene.

| Molecules/Atoms | Total Energy | Global Properties | | |
|---|---|---|---|---|
| | | $\chi$ | $\eta$ | $\omega$ |
| Sc | -760.62046 | 3.511 | 9.205 | 0.67 |
| Ti | -849.29000 | 3.648 | 4.195 | 1.587 |
| Fe | -1263.52195 | 4.265 | 4.415 | 2.06 |
| Ni | -1508.20025 | 4.524 | 4.044 | 2.53 |
| H | -0.50216 | 6.937 | 13.45 | 1.788 |
| H$_2$ | -1.17957 | 6.381 | 20.31 | 1.003 |
| C$_2$H$_4$ | -78.61551 | 4.422 | 12.34 | 0.792 |

values for gradual doping of molecular hydrogen (H$_2$) onto the C$_2$H$_4$---Sc$_n$ species initially increases for a single H$_2$ molecule but then shows a gradual decline for further trapping. The corresponding $\omega$ values increase for the first step and go on decreasing later. For

**Table 2:** Total energy of metal –ethylene and H₂ trapped metal-ethylene complexes computed at two different level of theory.

| Molecule | Energy B3LYP/6-311+g(d,p) | Energy MP2/6-311+g(d,p) |
|---|---|---|
| $C_2H_4$---Sc | -839.26605 | -838.09832 |
| $C_2H_4$---Sc---$H_2$ | -840.48256 | -839.29045 |
| $C_2H_4$----Sc----$2H_2$ | -841.63221 | -840.41415 |
| $C_2H_4$----Sc----$4H_2$ | -844.01601 | -842.75599 |
| $C_2H_4$----$Sc_2$ | -1599.91274 | -1597.85926 |
| $C_2H_4$----$Sc_2$---$2H_2$ | -1602.39254 | -1600.29399 |
| $C_2H_4$----$Sc_2$---$6H_2$ | -1607.12964 | -1604.95293 |
| $C_2H_4$----Ti | -927.93734 | -926.69325 |
| $C_2H_4$----Ti----$H_2$ | -929.22372 | -927.96802 |
| $C_2H_4$----Ti----$2H_2$ | -930.36065 | -929.08606 |
| $C_2H_4$----Ti----$4H_2$ | -932.76771 | -931.46154 |
| $C_2H_4$----Ti----$5H_2$ | -933.95478 | -932.63422 |
| $C_2H_4$-----$Ti_2$ | -1777.29820 | -1775.08555 |
| $C_2H_4$---$Ti_2$---$2H_2$ | -1779.78190 | -1777.52138 |
| $C_2H_4$---$Ti_2$---$6H_2$ | -1784.53331 | -1782.19569 |
| $C_2H_4$---$Ti_2$---$10H_2$ | -1789.27603 | -1786.90425 |
| $(C_2H_4$----$Ti_2)_2$ | -3554.73234 | -3550.42869 |
| $(C_2H_4$----$Ti_2)_2$-----$8H_2$ | -3564.32152 | -3559.87570 |
| $C_2H_4$----Fe | -1342.20082 | -1340.75577 |
| $C_2H_4$----Fe----$H_2$ | -1343.41044 | -1341.93243 |
| $C_2H_4$----Fe----$2H_2$ | -1344.62596 | -1343.19746 |
| $C_2H_4$----Fe----$3H_2$ | -1345.83724 | -1344.40370 |
| $C_2H_4$----Fe----$4H_2$ | -1347.03993 | -1345.60549 |
| $C_2H_4$----Ni | -1586.89990 | -1585.46814 |
| $C_2H_4$---Ni----$H_2$ | -1588.11078 | -1586.70740 |
| $C_2H_4$---Ni----$2H_2$ | -1589.31540 | -1587.89351 |
| $C_2H_4$-----$Ni_2$ | -3095.15522 | -3092.62685 |
| $C_2H_4$-----$Ni_2$---$2H_2$ | -3097.59183 | -3095.06366 |
| $C_2H_4$-----$Ni_2$---$4H_2$ | -3099.98537 | -3097.44174 |

the $C_2H_4$---$Sc_2$ system, nevertheless, an anomaly in the η and ω values for the hydrogen-doped systems are observed. A parallel intriguing trend of the variation of η and ω upon gradual hydrogen-trapping is also noticed among some of the systems in case of the corresponding $Ti_n$ – ethylene (n = 1, 2) clusters as envisaged from table 4. However, for the Fe and Ni-bound ethylene clusters the η

**Table 3**: Total energy (E, au), electronegativity ($\chi$, eV), hardness ($\eta$, eV) and electrophilicity ($\omega$, eV) of the Sc-ethylene system and its $H_2$-trapped analogs along with the atomic charge ($Q_k$) and Fukui functions for the metal site only.

| Molecules/Atoms | Global Properties | | | Local Properties | | |
|---|---|---|---|---|---|---|
| | $\chi$ | $\eta$ | $\omega$ | $Q_k$ | $f_k^+$ | $f_k^-$ |
| $C_2H_4$---Sc | 3.371 | 5.120 | 1.110 | 0.534 | 0.732 | 0.517 |
| $C_2H_4$---Sc---$H_2$ | 4.131 | 6.819 | 1.251 | 0.624 | 0.194 | 0.146 |
| $C_2H_4$---Sc---$2H_2$ | 3.238 | 5.742 | 0.913 | 0.574 | 1.204 | 0.186 |
| $C_2H_4$---Sc---$4H_2$ | 3.120 | 5.711 | 0.852 | 0.178 | 1.221 | 0.100 |
| $C_2H_4$---$Sc_2$ | 3.105 | 4.574 | 1.054 | 0.229 | 0.515 | 0.421 |
| | | | | 0.230 | 0.510 | 0.567 |
| $C_2H_4$---$Sc_2$--$2H_2$ | 4.485 | 7.492 | 1.342 | 0.673 | 0.243 | 0.048 |
| | | | | 0.673 | 0.243 | 0.048 |
| $C_2H_4$---$Sc_2$--$6H_2$ | 4.072 | 7.324 | 1.132 | 0.437 | 0.226 | -0.004 |
| | | | | 0.436 | 0.221 | -0.003 |

values are found to correlate pretty nicely with that of $\omega$ as relevant from tables 5 and 6 respectively. For the Fe and Ni-bonded complexes the hardness ($\eta$) increases uniformly with a gradual decrease in the consecutive ($\omega$) values thereby corroborating the associated principles of maximum hardness[37-39] and minimum electrophilicity[42, 43] which justify further stability of the clusters upon higher-order trapping. A glance at tables 7, 8, 9 and 10 for some plausible trapping reactions conceived between the metal – ethylene complexes $M_n$-($C_2H_4$) (M = Sc, Ti, Fe, Ni; n = 1, 2) and hydrogen

**Table 4**: Electronegativity ($\chi$, eV), hardness ($\eta$, eV) and electrophilicity ($\omega$, eV) of the Ti-ethylene system and its $H_2$-trapped analogs along with the atomic charge ($Q_k$) and Fukui functions for the metal site only.

| Molecules/Atoms | Global Properties | | | Local Properties | | |
|---|---|---|---|---|---|---|
| | $\chi$ | $\eta$ | $\omega$ | $Q_k$ | $f_k^+$ | $f_k^-$ |
| $C_2H_4$ – Ti | 3.477 | 2.891 | 2.090 | 0.538 | 0.768 | 0.408 |
| $C_2H_4$ – Ti -- $H_2$ | 4.599 | 7.797 | 1.356 | 0.516 | 0.288 | 0.100 |
| $C_2H_4$ – Ti -- $2H_2$ | 3.478 | 5.698 | 1.061 | 0.266 | 0.381 | 0.185 |
| $C_2H_4$ – Ti -- $4H_2$ | 3.100 | 7.066 | 0.680 | -0.015 | 0.149 | -0.072 |
| $C_2H_4$ – Ti -- $5H_2$ | 3.023 | 6.695 | 0.683 | -0.326 | 0.051 | -0.050 |
| $C_2H_4$--$Ti_2$ | 2.812 | 3.424 | 1.155 | 0.220 | 0.625 | 0.570 |
| | | | | 0.220 | 0.626 | 0.514 |
| $C_2H_4$--$Ti_2$--$2H2$ | 3.771 | 4.317 | 1.647 | 0.484 | 0.257 | 0.104 |
| | | | | 0.483 | 0.336 | 0.105 |
| $C_2H_4$--$Ti_2$--$6H2$ | 3.752 | 3.384 | 2.080 | 0.114 | 0.082 | -3E-05 |
| | | | | -0.119 | 0.157 | 0.158 |
| $C_2H_4$--$Ti_2$--$10H_2$ | 2.957 | 4.739 | 0.922 | -0.548 | -0.050 | 0.057 |
| | | | | -0.304 | -0.050 | -0.032 |
| ($C_2H_4$ --- Ti)$_2$ | 2.034 | 2.047 | 1.010 | 0.063 | 0.075 | 0.170 |
| | | | | 0.317 | 0.644 | 0.353 |
| | | | | 0.063 | 0.075 | 0.170 |
| | | | | 0.317 | 0.644 | 0.353 |
| ($C_2H_4$ --- Ti)$_2$ --- $8H_2$ | 2.595 | 1.300 | 2.590 | 0.020 | 0.031 | 0.146 |
| | | | | -0.082 | 0.173 | 0.092 |
| | | | | 0.020 | -0.010 | 0.147 |
| | | | | -0.082 | 0.173 | 0.093 |

molecule for the metals Sc, Ti, Fe and Ni respectively reveals that for all the reactions, the corresponding interaction energies (IE) and reaction electrophilicities ($\Delta\omega$) are negative. This is quite enthusing as a negative IE or $\Delta\omega$ value for a chemical process affirms considerable thermodynamic feasibility to the same, thereby, rendering ample stability to the resultant $H_2$-trapped complexes as well. The average dissociative chemisorption energies ($\Delta E_{CE}$)

**Table 5**: Electronegativity ($\chi$, eV), hardness ($\eta$, eV) and electrophilicity ($\omega$, eV) of the Fe-ethylene system and its $H_2$-trapped analogs along with the atomic charge ($Q_k$) and Fukui functions for the metal site only.

| Molecules/Atoms | Global Properties | | | Local Properties | | |
|---|---|---|---|---|---|---|
| | $\chi$ | $\eta$ | $\omega$ | $Q_k$ | $f_k^+$ | $f_k^-$ |
| $C_2H_4$ ---- Fe | 5.092 | 6.103 | 2.124 | 0.026 | 0.528 | 0.634 |
| $C_2H_4$ ---- Fe --- $H_2$ | 4.916 | 6.223 | 1.942 | 0.057 | 0.204 | 0.223 |
| $C_2H_4$ ---- Fe --- $2H_2$ | 4.477 | 6.596 | 1.520 | -0.108 | 0.366 | 0.225 |
| $C_2H_4$ ---- Fe --- $3H_2$ | 4.265 | 7.249 | 1.254 | -0.339 | 0.242 | 0.097 |
| $C_2H_4$ ---- Fe --- $4H_2$ | 3.794 | 8.211 | 0.877 | -0.528 | 1.089 | 0.028 |

computed for the trapping of hydrogen molecule onto the metal – ethylene system is positive which favors the incoming $H_2$ species to bind with the metal site with formation of new bonds. However, with increasing number of hydrogen atoms being adsorbed, the

**Table 6**: Electronegativity ($\chi$, eV), hardness ($\eta$, eV) and electrophilicity ($\omega$, eV) of the Ni-ethylene system and its $H_2$-trapped analogs along with the atomic charge ($Q_k$) and Fukui functions for the metal site only.

| Molecules/Atoms | Global Properties | | | Local Properties | | |
|---|---|---|---|---|---|---|
| | $\chi$ | $\eta$ | $\omega$ | $Q_k$ | $f_k^+$ | $f_k^-$ |
| $C_2H_4$ ---- Ni | 3.676 | 6.842 | 0.987 | 0.097 | 1.019 | 0.529 |
| $C_2H_4$ ---- Ni --- $H_2$ | 3.616 | 7.233 | 0.904 | -0.027 | 3E-04 | 0.365 |
| $C_2H_4$ ---- Ni --- $2H_2$ | 3.558 | 8.078 | 0.784 | -0.120 | 1.644 | 0.266 |
| $C_2H_4$ ---- $Ni_2$ | 3.546 | 5.608 | 1.121 | 0.051 | 0.537 | 0.409 |
| | | | | 0.052 | 0.533 | 0.411 |
| $C_2H_4$ ---- $Ni_2$ --- $2H_2$ | 2.779 | 5.65 | 0.683 | -0.009 | 0.293 | 0.524 |
| | | | | -0.009 | 0.293 | 0.537 |
| $C_2H_4$ ---- $Ni_2$ --- $4H_2$ | 3.185 | 7.563 | 0.671 | -0.128 | 0.649 | 0.198 |
| | | | | -0.127 | 0.648 | 0.195 |

$\Delta E_{CE}$ values decrease for all the metal – ethylene $M_n$-$(C_2H_4)$ (M = Sc, Ti, Fe, Ni; n = 1, 2) molecules. This reveals that with an increase in H-crowding around the metal – ethylene core, the driving force for $H_2$ dissociation may further go on a decline.[46] A careful scrutiny of figure 1 shows that for the Sc-bound ethylene complexes, the hydrogen atoms are linked to the central metal atom in both the atomic and molecular forms. An in-depth study of the H – H bond distances from table 11 and the corresponding atomic charges on the H-centers of the trapped $H_2$ molecules from figure 1 unveils something really unique which further gives a better impetus towards understanding the driving force responsible for the possible feasibility of hydrogen storage in such metal – ethylene systems. From figure 1 it is quite transparent that the H$_2$ moieties bound to the central metal site mostly in its molecular form bear a slight positive charge on the respective H-centers. The H – H bond length, however, increases than that of its original molecular form thereby rendering the H$_2$ system to behave nonetheless as a ligand. The increment in the H – H distance upon complexation (rather trapping) can be compared with an analogous increase in the C = C bond length in C$_2$H$_4$ that has already been explained in literature on the backdrop of the Dewar-Chatt-Duncanson (DCD) model of molecular binding. A similar behavior exhibited by the incoming $H_2$ molecule can be attributed to the Kubas-model[21] of metal – dihydrogen interaction. The $H_2$ molecule acting as a $\eta^2$- ligand system is bonded to the transition metal and thus stabilized. However, the $H_2$ molecules that are intended to be bound to the transition metal site mostly in its atomic form are found to split wide-apart with large H – H distances as compared to the free, unbound molecular form. The atomic charges on the corresponding sites of the lone hydrogen atoms are noticeably high compared to that of the $\eta^2$-$H_2$ ligands. Thus the interaction between the central transition metal and the atomic hydrogens are supposed to be fairly electrostatic in nature. So, the chemical binding of molecular hydrogen onto the metal – ethylene moiety to form stable $H_2$-trapped analogues have got two facets – a Kubas-type binding between the metal center and the $\eta^2$-H$_2$ ligands which further sustain their molecular nature in the trapped form, and, an electrostatic model of binding between the comparably highly charged atomic hydrogens (split wide-open) and the central metal atom. The important frontier molecular orbitals (FMOs) of some representative metal – ethylene complexes displayed in figure 5 reveals the dominance of a σ-antibonding character in most of the HOMO contours. Nevertheless, the presence of a bulge of electron density near the metal sites provides ample evidence of chemical binding between the metal center and ethylene moiety. The stability of the M - $\eta^2$-H$_2$ linkage might be due to some favorable interactions between the metal-*d*-orbitals and the σ-antibonding orbitals of hydrogen. The overall electron density distribution pattern in the frontier orbitals of different hydrogen-trapped complexes reveals that the electrons are delocalized over the entire molecular skeleton.

**Table 7**: Plausible $H_2$-trapping reactions for Sc-ethylene complexes showing their interaction energy (IE), interaction energy per $H_2$ molecule (IE/$H_2$), reaction electrophilicity ($\Delta\omega$) and dissociative chemisorption energy ($\Delta E_{CE}$)

| Reactions | Reaction Energetics | | | |
|---|---|---|---|---|
| | IE | IE/$H_2$ | $\Delta\omega$ | CE |
| $C_2H_4$---Sc + $H_2$ = $C_2H_4$---Sc---$H_2$ | -23.18 | -23.18 | -0.861 | 23.18 |
| $C_2H_4$---Sc + $2H_2$ = $C_2H_4$---Sc---$2H_2$ | -4.40 | -2.20 | -2.202 | 2.20 |
| $C_2H_4$---Sc + $4H_2$ = $C_2H_4$---Sc---$4H_2$ | -19.88 | -4.97 | -4.268 | 4.97 |
| $C_2H_4$---$Sc_2$ + $2H_2$ = $C_2H_4$---$Sc_2$---$2H_2$ | -75.71 | -37.90 | -1.717 | 37.86 |
| $C_2H_4$---$Sc_2$ + $6H_2$ = $C_2H_4$---$Sc_2$---$6H_2$ | -87.52 | -14.60 | -5.937 | 14.59 |

**Table 8**: Plausible H$_2$-trapping reactions for Ti-ethylene complexes showing their interaction energy (IE), interaction energy per H$_2$ molecule (IE/H$_2$), reaction electrophilicity (Δω) and dissociative chemisorption energy (ΔE$_{CE}$)

| Reactions | Reaction Energetics | | | |
|---|---|---|---|---|
| | IE | IE/H$_2$ | Δω | CE |
| C$_2$H$_4$---Ti + H$_2$ = C$_2$H$_4$---Ti---H$_2$ | -67.03 | -67.03 | -1.737 | 67.03 |
| C$_2$H$_4$---Ti + 2H$_2$ = C$_2$H$_4$---Ti---2H$_2$ | -40.27 | -20.13 | -3.034 | 20.13 |
| C$_2$H$_4$---Ti + 4H$_2$ = C$_2$H$_4$---Ti---4H$_2$ | -70.34 | -17.59 | -5.420 | 17.59 |
| C$_2$H$_4$---Ti + 5H$_2$ = C$_2$H$_4$---Ti---5H$_2$ | -75.04 | -15.01 | -6.420 | 15.01 |
| C$_2$H$_4$---Ti$_2$ + 2H$_2$ = C$_2$H$_4$---Ti$_2$---2H$_2$ | -78.16 | -39.08 | -1.513 | 39.08 |
| C$_2$H$_4$---Ti$_2$ + 6H$_2$ = C$_2$H$_4$---Ti$_2$---6H$_2$ | -98.95 | -16.49 | -5.091 | 16.49 |
| C$_2$H$_4$---Ti$_2$ + 10H$_2$ = C$_2$H$_4$---Ti$_2$---10H$_2$ | -114.30 | -11.43 | -10.26 | 11.43 |
| (C$_2$H$_4$---Ti$_2$)$_2$ + 8H$_2$ = (C$_2$H$_4$---Ti$_2$)$_2$ ---8H$_2$ | -95.77 | -11.97 | -6.440 | 11.97 |

**Table 9**: Plausible $H_2$-trapping reactions for Fe-ethylene complexes showing their interaction energy (IE), interaction energy per $H_2$ molecule (IE/$H_2$), reaction electrophilicity ($\Delta\omega$) and dissociative chemisorption energy ($\Delta E_{CE}$)

| Reactions | Reaction Energetics | | | |
|---|---|---|---|---|
| | IE | IE/$H_2$ | $\Delta\omega$ | CE |
| $C_2H_4$---Fe + $H_2$ = $C_2H_4$---Fe---$H_2$ | -18.85 | -18.85 | -1.185 | 18.85 |
| $C_2H_4$---Fe + $2H_2$ = $C_2H_4$---Fe---$2H_2$ | -41.41 | -20.71 | -2.610 | 20.71 |
| $C_2H_4$---Fe + $3H_2$ = $C_2H_4$---Fe---$3H_2$ | -61.31 | -20.44 | -2.875 | 20.44 |
| $C_2H_4$---Fe + $4H_2$ = $C_2H_4$---Fe---$4H_2$ | -75.82 | -18.96 | -3.253 | 18.96 |

**Table 10**: Plausible $H_2$-trapping reactions for Ni-ethylene complexes showing their interaction energy (IE), interaction energy per $H_2$ molecule (IE/$H_2$), reaction electrophilicity ($\Delta\omega$) and dissociative chemisorption energy ($\Delta E_{CE}$)

| Reactions | Reaction Energetics | | | |
|---|---|---|---|---|
| | IE | IE/$H_2$ | $\Delta\omega$ | CE |
| $C_2H_4$---Ni + $H_2$ = $C_2H_4$---Ni---$H_2$ | -19.65 | -19.65 | -1.086 | 19.65 |
| $C_2H_4$---Ni + $2H_2$ = $C_2H_4$---Ni---$2H_2$ | -35.36 | -17.68 | -2.209 | 17.68 |
| $C_2H_4$---$Ni_2$ + $2H_2$ = $C_2H_4$---$Ni_2$---$2H_2$ | -48.62 | -24.31 | -2.443 | 24.31 |
| $C_2H_4$---$Ni_2$ + $4H_2$ = $C_2H_4$---$Ni_2$---$4H_2$ | -70.20 | -17.55 | -4.460 | 17.55 |

**Table 11**: The molecular point groups (PG) and the H – H bond distances (Å) of the $H_2$ ligands trapped in different forms (atomic and molecular) onto the metal – ethylene complexes.

| Complexes | Point Group (PG) | H-H Distance (Molecular form) | H-H Distance (atomic form) |
|---|---|---|---|
| $C_2H_4$---Sc---$H_2$ | $C_1$ | | 3.22 |
| $C_2H_4$---Sc---$2H_2$ | $C_1$ | 0.80, 0.80 | |
| $C_2H_4$---Sc---$4H_2$ | $C_1$ | 0.80, 0.80, 0.80, 0.80 | |
| $C_2H_4$---$Sc_2$--$2H_2$ | $C_1$ | | 3.19, 3.19 |
| $C_2H_4$---$Sc_2$--$6H_2$ | $C_1$ | 0.76, 0.76, 0.76, 0.76 | 3.27, 3.28 |
| $C_2H_4$ – Ti -- $H_2$ | $C_1$ | | 3.04 |
| $C_2H_4$ – Ti -- $2H_2$ | $C_1$ | 0.87, 0.87 | |
| $C_2H_4$ – Ti -- $4H_2$ | $C_2$ | 0.82, 0.82, 0.82, 0.82 | |
| $C_2H_4$ – Ti -- $5H_2$ | $C_S$ | 0.77, 0.82, 0.82, 0.82, 0.82 | |
| $C_2H_4$--$Ti_2$--$2H_2$ | $C_{2h}$ | | 3.05, 3.05 |
| $C_2H_4$--$Ti_2$--$6H_2$ | $C_1$ | 0.82, 0.82, 0.77, 0.77 | 3.13, 2.99 |
| $C_2H_4$--$Ti_2$--$10H_2$ | $C_S$ | 0.84, 0.87, 0.89, 0.85, 0.84, 0.82, 0.82, 0.82, 0.76, 0.82 | |
| $(C_2H_4$ --- $Ti)_2$ --- $8H_2$ | $C_1$ | 0.82, 0.82, 0.82, 0.82, 0.82, 0.82, 0.82, 0.82 | |
| $C_2H_4$ ---- Fe --- $H_2$ | $C_S$ | | 1.96 |
| $C_2H_4$ ---- Fe --- $2H_2$ | $C_1$ | 0.88, 0.88 | |
| $C_2H_4$ ---- Fe--- -$3H_2$ | $C_1$ | 0.83, 0.88, 0.87 | |
| $C_2H_4$ ---- Fe --- $4H_2$ | $C_S$ | 0.81, 0.88, 0.81, 0.89 | |
| $C_2H_4$ ---- Ni --- $H_2$ | $C_1$ | 0.83 | |
| $C_2H_4$ ---- Ni --- $2H_2$ | $C_{2V}$ | 0.83, 0.83 | |
| $C_2H_4$ ---- $Ni_2$ --- $2H_2$ | $C_1$ | 0.85, 0.85 | |
| $C_2H_4$ ---- $Ni_2$ --- $4H_2$ | $D_{2h}$ | 0.84, 0.84, 0.84, 0.84 | |

Tables 3 – 6 also demonstrate that the Mulliken charges computed for the central metal sites show an unusual variation upon gradual complexation with the incoming $H_2$ molecules. The metal atoms that are known to be an electropositive species are found to bear a negative charge upon increasing $H_2$ trapping. It has been observed by other researchers as well.[47] It is assumed to be due to the drawback of the Mulliken as well as the natural population analysis schemes. In this study it is also observed that a negative charge is gradually piled up onto the metal atom with gradual loading of hydrogen. Again, most of the incoming $H_2$ molecules are bonded to the metal atom in the Kubas fashion augmenting the same obtained through the back donation from ethylene as envisaged from figures 1 – 4 which prompts us to suggest a rigorous charge transfer that occurs from the $\eta^2$-$H_2$ ligands to the metal atom which allows the latter to sustain a formal negative charge. The cumulative effect of charge donation by a number of $H_2$-ligands in the Kubas fashion renders the metal center formally negative. The bulge of electron density around the metal sites as evident from the frontier orbitals also indirectly justifies the fact.

## Conclusion

A comprehensive study of the binding of hydrogen onto some metal – ethylene complexes, $M_n$-$(C_2H_4)$ (M = Sc, Ti, Fe, Ni; n = 1, 2) have been performed using conceptual DFT based reactivity descriptors. The efficiency of such metal complexes towards trapping of molecular hydrogen is quite unequivocal. The stability of the trapped complexes has been understood both from the viewpoint of increasing hardness (η) trends and a subsequent decrease in the electrophilicity (ω) in most cases upon gradual hydrogen loading as well as on the basis of the Kubas-model. The ability of transition metals other than Sc or Ti to act as potential storage stuffs for hydrogen in the presence of a π- system (like ethylene) has been demonstrated. Presumably the existence of a negative charge on the metal sites upon gradual $H_2$-binding stems from an increasingly favorable orbital interaction between the incoming $\eta^2$- dihydrogen ligands and the central metal atom. The binding mode of molecular hydrogen to the metal site through this Kubas-type interaction is established. Nonetheless, the usage of other transition metals associated with π-systems other than ethylene that may serve as effective materials for hydrogen storage has opened an area that deserves to be widely cultivated. Further work is in progress.

## Acknowledgements

We thank Indo- EU (HYPOMAP) project for financial assistance. One of the authors (AC) would like to acknowledge the CTS, IIT Kharagpur for a Visitors' Fellowship.